\documentclass[12pt]{article}

\usepackage{amssymb,amsmath}

\usepackage{graphicx}
\usepackage{subfigure}

\newcommand{\be}{\begin{equation}}
\newcommand{\ee}{\end{equation}}
\newcommand{\Dlt}{\Delta}

\newcommand{\prt}{\partial}
\newcommand{\br}{{\bf r}}
\newcommand{\bk}{{\bf k}}

\newcommand{\bt}{\beta}

\newcommand{\ra}{\rightarrow}

\newcommand{\Om}{\Omega}

\newcommand{\lbd}{\lambda}

\newcommand{\cF}{{\cal F}}
\newcommand{\cH}{{\cal H}}
\newcommand{\cM}{{\cal M}}
\newcommand{\cD}{{\cal D}}

\newcommand{\lgl}{\langle}
\newcommand{\rgl}{\rangle}

\begin{document}

\begin{center}

{\Large{\bf Models of mixed hadron-quark-gluon matter} \\ [5mm]

V.I. Yukalov$^1$ and E.P. Yukalova$^2$ } \\ [3mm]

{\it $^1$Bogolubov Laboratory of Theoretical Physics, \\
Joint Institute for Nuclear Research, Dubna 141980, Russia \\ [3mm]

$^2$Laboratory of Information Technologies, \\
Joint Institute for Nuclear Research, Dubna 141980, Russia \\ [3mm] }

\end{center}

\vskip 3cm

\begin{abstract}
The problem of the possible creation of mixed hadron-quark-gluon matter,
that can arise at nuclear or heavy-ion collisions, is addressed. It is
shown that there can exist several different kinds of such a mixed matter.
The main types of this matter can be classified onto macroscopic mixture,
mesoscopic mixture, and microscopic mixture. Different types of these
mixtures require principally different descriptions. Before comparing
theoretical results with experiments, one has to analyze thermodynamic
stability of all these mixed states, classifying them onto unstable,
metastable, and stable. Only the most stable mixed state should be
compared with experiment. Mixed states also need to be checked with
regard to stratification instability. In addition to the static
stratification instability, there can happen dynamic instability
occurring in a mixture of components moving with respect to each other.
This effect, called counterflow instability, has also to be taken into
account, since it can lead to the stratification of mixed matter.
\end{abstract}

\newpage

\section{Introduction}

At high temperatures and/or densities, hadronic matter is expected to
undergo a transition to quark-gluon plasma, where quarks and gluons are no
longer confined inside hadrons but can propagate much further in extent
than the typical sizes of hadrons. Such a deconfinement transition can
happen under heavy-ion or nuclear collisions. It is assumed to exist
in the early universe cosmology, since for a time on the order of the
microsecond the temperature was high enough for the elementary degrees of
freedom of QCD to be in a deconfined state. The quark-gluon plasma can
also exist in the interior of compact stars.

The peculiarities of the transition from hadronic matter to quark-gluon
plasma, that is, of the deconfinement transition, have been the object of
many discussions (see review articles [1-10]). From general arguments, it is
impossible to infer the order of the QCD transition, whether it is
1-st order, 2-nd order, or crossover. Being based on a model consideration,
the deconfinement was shown to be a gradual crossover [6,7]. At the present
time, this result has been confirmed by numerical simulations of lattice
QCD showing convincingly that deconfinement is really a crossover [11-15].

In order to be able to describe the states of matter and phase transitions
in thermodynamic terms, it is required that the matter be at least in
quasi-equilibrium. The experimental lifetime of fireballs, formed under
heavy-ion collisions, is of order $t_{exp} \sim 10^{-22}$ s [16,17]. The
local equilibration time of nuclear matter is $t_{loc} \sim 10^{-23}$ s
[18,19]. Since $t_{loc} \ll t_{exp}$, equilibration is feasible and
thermodynamic language is applicable to treating the fireball states.

A plausible assumption is that in the process of the transformation of
hadronic matter into quark-gluon plasma there can arise an intermediate
state of matter representing a mixture of hadronic and quark-gluon states
[20,21]. Note that the manifestation of quark degrees of freedom, resulting
in the appearance of the Blokhintsev fluctons [22], Baldin cumulative
effect [23], and in the formation of multi-quark clusters, has also
been assumed to occur even at temperatures essentially lower than the
deconfinement point [24-31].

However, the nature of the mixed hadron-quark-gluon state has not been
well understood. It is the aim of the present paper to explain that,
actually, there can exist several kinds of such a mixed state, with the
main three types that can be classified onto macroscopic, mesoscopic, and
microscopic mixed states. These states have rather different properties
and require essentially different theoretical description.

In the paper, we use the system of units, where the Planck and Boltzmann
constants are set to one.

\section{Macroscopic mixed state}

This type of mixed state would arise if the deconfinement transition would
be of first order [32-34]. Then, at the phase-transition point, the system,
say fireball of a linear size $L$, separates into macroscopic domains of
size $l$ corresponding to hadron phase and quark-gluon phase, so that
\be
\label{1}
a \ll l \sim L \;   ,
\ee
where $a$ is mean interparticle distance. The domains are macroscopic,
being of order of the system size $L$. They also are called droplets or
blobs, or bubbles [35-40]. Their topology is similar to the droplets of
nucleons arising in the low-density nuclear matter [41,42]. The domains
of different phases correspond to different vacua [43-46], with the
physical states of different domains being mutually orthogonal [47].

Strictly speaking, under first-order phase transition, mixed phase can
occur only at the transition point, where two kinds of pure phases meet
each other, pure hadron phase and pure quark-gluon phase. The qualitative
behavior of the transition temperature $T_c$ as a function of barion
density $\rho_B$ is shown in Fig. 1. Hadron matter consists of only
hadrons, interacting with each other through hadron-hadron interactions
[48,49]. Pure quark-gluon plasma is described by an equation of state for
free quarks and gluons, with taking into account their interactions [50]
by incorporating some non-perturbative effects [51-53].

The straightforward order parameters are the density of hadron matter,
$\rho_h$ and the density of quark-gluon plasma, $\rho_q$. In the hadronic
matter
\be
\label{2}
  \rho_h > 0 \; , \qquad \rho_q \equiv 0 \; ,
\ee
while in the quark-gluon plasma
\be
\label{3}
 \rho_h \equiv 0 \; , \qquad \rho_q > 0 \;  .
\ee
It is also possible to use as an order parameter the Wilson loop [54,55].

Each type of particles is characterized by barion number $B_i$ and
strangeness $S_i$. For simplicity, the particles are assumed to be neutral.
The chemical potential of the $i$-type particles is expressed as
\be
\label{4}
\mu_i = \mu_B B_i + \mu_S S_i
\ee
through the barion, $\mu_B$, and strangeness, $\mu_S$ chemical potentials.
The barion and strangeness densities are given by the relations
\be
\label{5}
 \rho_B = \sum_i B_i \rho_i = \frac{\prt P}{\prt\mu_B} \;  , \qquad
\rho_S = \sum_i S_i \rho_i = \frac{\prt P}{\prt\mu_S} \;  ,
\ee
in which $P$ is pressure and $\rho_i$ is the density of the $i$-type
particles. The behavior of pressure, under first-order phase transition,
is shown in Fig. 2. Since the grand potential is $\Omega = - PV$, where
$V$ is the system volume, the larger pressure corresponds to the lower
grand potential.

The transition temperature is defined by the equality of the pressures,
\be
\label{6}
 P_h(T_c,\mu_B) = P_q(T_c,\mu_B) \;  ,
\ee
where, for simplicity, the strangeness density is fixed. This gives
$T_c = T_c(\mu_B)$. Because from the left and the right of $T_c$, the
pressures are different, we have two barion densities, for hadrons and
for plasma,
\be
\label{7}
  \rho_{Bh} = \frac{\prt P_h}{\prt\mu_B} =  \rho_{Bh}(T,\mu_B) \; ,
\qquad
\rho_{Bq} = \frac{\prt P_q}{\prt\mu_B} =  \rho_{Bq}(T,\mu_B) \; ,
\ee
which gives two barion potentials, $\mu_{Bh}$ and $\mu_{Bq}$ that coincide
at the transition temperature:
\be
\label{8}
 \mu_{Bh}(T_c,\rho_{Bh}) = \mu_{Bq}(T_c,\rho_{Bq}) \;  .
\ee
This defines $T_c = T_c(\rho_{Bh}, \rho_{Bp})$.

The point of a first-order phase transition is the point of instability.
Infinitesimally small fluctuations of temperature around $T_c$ will result
in finite jumps between two different barion densities in Eq. (\ref{7}). So that
the mixed phase at this point is unstable.

One says that the mixed phase could exist not merely at the transition
point, but also in a region around it. This is explained as being due to
the Maxwell construction that is demonstrated in Fig. 3 for the pressure
as a function of the reduced barion volume
$$
 v_B \equiv \frac{1}{\rho_B} \;  .
$$
Here, the standard behavior of the pressure under a first-order phase
transition [56] is corrected by replacing the part, corresponding to unstable
and metastable states (shown by the dashed line), by the horizontal solid
line between the barion volumes
$$
 v_{Bh} \equiv \frac{1}{\rho_{Bh} } \; , \qquad
 v_{Bq} \equiv \frac{1}{\rho_{Bq} } \;  .
$$

As a result of this construction, the phase diagram of Fig. 1 transforms
into that of Fig. 4. The Maxwell construction for the pressure as a function
of temperature is equivalent to the smoothing of the pressure, as is shown
in Fig. 5. The mixed phase exists between the low, $T_n$, and upper, $T_n^*$,
nucleation temperatures.

However, as is evident from Fig. 3, on the coexistence line, one has
\be
\label{9}
 \frac{\prt P}{\prt v_B} = 0 \qquad ( v_{Bh} < v_B < v_{Bq} ) \; .
\ee
This implies that the isothermal compressibility
$$
\kappa_T = - \; \frac{1}{v_B} \left ( \frac{\prt P}{\prt v_B}
\right )^{-1}
$$
is divergent everywhere in the region of the mixed phase existence:
\be
\label{10}
\kappa_T \ra \infty \qquad ( T_n < T < T_n^* ) \;   .
\ee
The divergence of the compressibility means instability, since
infinitesimally weak pressure fluctuations would lead to the system explosion
during the short explosion time
$$
 t_{exp} \sim \frac{1}{\kappa_T} \;  .
$$
That is, the fireball would explode even before it could equilibrate.

Concluding, if deconfinement would be a first-order phase transition, then,
formally, a mixed hadron-quark-gluon phase could arise around the transition
point, however such a mixed state is strongly unstable and, in reality,
cannot exist as an equilibrium phase. In addition, as QCD lattice simulations
prove [11-15], deconfinement is not a first-order transition, but rather a
crossover.

\section{Mesoscopic mixed state}

There can exist another type of mixed state that can be called mesoscopic
mixed hadron-quark-gluon matter. The term "mesoscopic" means that the typical
size $l_g$ of the arising germs of one phase inside the other is between the
mean interparticle distance and the system size:
\be
\label{11}
  a \ll l_g \ll L \; .
\ee
Below the deconfinement temperature, these are the germs of quark-gluon
plasma surrounded by hadron matter. And above the transition temperature,
these are the germs of hadron matter inside quark-gluon plasma.

For the mesoscopic mixed state, pressure is uniquely defined and does not
require the phase transition to be of the first order. Generally, it can
be of any order, including the crossover type [19,57]. Mesoscopic mixed
state can exist in a large temperature interval between the low and upper
nucleation temperatures.

The mesoscopic mixed state is basically different from the macroscopic one,
exhibiting the following main features.

(i) The germs of competing phases do not need to be in absolute equilibrium.
They can have finite lifetime $t_g$. But they are to be in quasi-equilibrium,
such that the local equilibration time be essentially shorter than their
lifetime:
\be
\label{12}
 t_{loc} \ll t_g \;  .
\ee

(ii) The spatial distribution of germs at a snapshot is random. They form no
ordered spatial structure, such as domains.

(iii) The spatial distribution of germs is also random with respect to
repeated experiments.

(iv) The typical size of the germs is mesoscopic in the sense of Eq. (\ref{11}).

(v) The germ geometry is of multiscale nature. Their shapes are not regular,
but are rather ramified. And the sizes $l_g$ lie in a dense interval
$[l_g^{min}, l_g^{max}]$, such that
$$
 a \ll l_g^{min} < l_g^{max} \ll L \;  .
$$

The description of the mesoscopic mixed state has to take into account the
generic random nature of the spatial germ distribution. The main ideas of the
theory are as follows [19,57,58]. We keep in mind a mixture of two phases,
e.g., hadron matter and quark-gluon plasma.

At a snapshot, the system volume is divided onto the volumes of different
phases,
$$
 \mathbb{V} = \mathbb{V}_1 \bigcup \mathbb{V}_2 \;  ,
$$
separated by the Gibbs equimolecular separating surface, for which extensive
observable quantities are additive. This concerns as well the number of
particles in each phase and the related volumes:
\be
\label{13}
 N = N_1 + N_2 \; , \qquad V = V_1 + V_2 \; ,
\ee
with $V_\nu \equiv {\rm mes}\mathbb {V}_\nu$. Mathematically, the separation
is characterized by the manifold indicator functions
\begin{eqnarray}
\label{14}
\xi_\nu(\br) = \left \{ \begin{array}{ll}
1 , & ~~ \br \in \mathbb{V}_\nu \\
0 , & ~~ \br \not\in \mathbb{V}_\nu \; ,
\end{array} \right.
\end{eqnarray}
where ${\bf r}$ is a spatial variable and $\nu = 1,2$ enumerates the phases.

At a snapshot, the mixture needs to be described by a representative
statistical ensemble $\{\mathcal{H}, \hat{\rho}(\xi)\}$, where $\mathcal{H}$
is the space of microstates and $\hat{\rho}(\xi)$ is a statistical
operator [58]. The space of microstates is given by the fiber space
\be
\label{15}
 \cH = \cH_1 \bigotimes \cH_2 \;  ,
\ee
with the fiber bases $\mathcal{H}_\nu$ being weighted Hilbert spaces. The
statistical operator is normaized as
\be
\label{16}
 {\rm Tr} \int \hat\rho(\xi)\; \cD\xi = 1 \;  ,
\ee
by taking the trace over the quantum degrees of freedom and averaging over
the random germ spatial configurations defined through the functional
integral over the manifold indicator functions (\ref{14}).

To construct a representative ensemble, one defines the internal energy
\be
\label{17}
 E =  {\rm Tr} \int \hat\rho(\xi)\hat H(\xi)\; \cD\xi
\ee
and all constraining quantities
\be
\label{18}
 C_i =  {\rm Tr} \int \hat\rho(\xi)\hat C_i(\xi)\; \cD\xi \; ,
\ee
required for the unique description of the system. The statistical operator
is found from the principle of minimal information, by minimizing the
information functional
$$
I [ \hat\rho(\xi) ] = {\rm Tr} \int \hat\rho(\xi)
\ln \hat\rho(\xi)\; \cD\xi \; +
$$
$$
+ \; \lbd_0 \left [
{\rm Tr} \int \hat\rho(\xi)\; \cD\xi - 1 \right ] \; +
\bt  \left [
{\rm Tr} \int \hat\rho(\xi) \hat H(\xi) \; \cD\xi - E \right ] \; +
$$
\be
\label{19}
+ \; \sum_i \lbd_i \left [
{\rm Tr} \int \hat\rho(\xi)\hat C_i(\xi)\; \cD\xi - C_i \right ] \; ,
\ee
in which $\lambda_0$, $\beta$, and $\lambda_i$ are Lagrange multipliers.

The minimization yields the statistical operator
\be
\label{20}
 \hat\rho(\xi) = \frac{1}{Z} \; \exp \{ - \bt H(\xi) \} \;  ,
\ee
with the grand Hamiltonian
\be
\label{21}
  H(\xi) = \hat H(\xi) - \sum_i \mu_i \hat C_i(\xi) \; ,
\ee
where $\mu_i \equiv - \lambda_i T$. The partition function is
$$
Z =  {\rm Tr} \int \exp\{-\bt H(\xi) \} \; \cD\xi \;  ,
$$
and $\beta = 1/T$ is inverse temperature.

Let us introduce the effective Hamiltonian $\tilde{H}$ defined by the
equality
\be
\label{22}
  \int \exp\{-\bt H(\xi) \} \; \cD\xi  = \exp(-\bt\widetilde H) \; .
\ee
After this, the partition function reduces to the form
$$
 Z = {\rm Tr} e^{-\bt\widetilde H} \;  ,
$$
containing only the trace over quantum degrees of freedom.

The geometric weights of each phase are given by the expressions
\be
\label{23}
 w_\nu = \int \xi_\nu(\br) \; \cD\xi  ,
\ee
satisfying the normalization condition
\be
\label{24}
  w_1 + w_2 = 1 \; .
\ee
This provides the minimum for the grand potential
$$
 \Om = - T \ln {\rm Tr} e^{-\bt\widetilde H} \;  ,
$$
that can be found from the conditions
\be
\label{25}
 \frac{\prt\Om}{\prt w_\nu} = 0 \; \qquad
 \frac{\prt^2\Om}{\prt w_\nu^2}  > 0 \;,
\ee
taking into account normalization (\ref{24}). The phase weights (\ref{23})
play the role of additional order parameters characterizing the mixed
state [19,57,60,61].

The mesoscopic mixed state is stable, with deconfinement being rather
a sharp crossover.

\section{Microscopic mixed state}

The third type of mixture is termed microscopic because hadrons are
uniformly intermixed with quark-gluon plasma, without forming either germs
or droplets. Such a mixed state can be treated by the theory of
clustering matter [6,7], considering hadrons as quark clusters. Each
kind of clusters, enumerated by the index $i$, is characterized by
the barion number $B_i$, strangeness $S_i$, and compositeness $z_i$.
The latter shows the number of quarks forming a cluster of that type.
For instance, the quark compositeness is 1, meson compositeness is 2,
and the nucleon compositeness is 3.

The space of microstates for the mixture is the tensor product
\be
\label{26}
 \cM = \bigotimes_i \cF_i \;  ,
\ee
in which $\mathcal{F}_i$ is the Fock space for the $i$-clusters.

The density of $i$-clusters is
\be
\label{27}
  \rho_i = \zeta_i \int n_i(\bk) \; \frac{d\bk}{(2\pi)^3} \; ,
\ee
where $\zeta_i$ is a degeneracy factor and $n_i({\bf k})$ is a
momentum distribution. The total mean quark density is
\be
\label{28}
 \rho = \sum_i z_i \rho_i \;  .
\ee
The cluster weights are defined by the ratio
\be
\label{29}
 w_i \equiv \frac{z_i\rho_i}{\rho} \;  ,
\ee
which gives $w_i = w_i(\rho_B, \rho_S, T)$. By definition, one has
$$
 0 \leq w_i \leq 1 \; , \qquad \sum_i w_i = 1 \;  .
$$

The Hamiltonian of a microscopic mixture, generally, has the form
\be
\label{30}
  \hat H = \sum_i \hat H_i +
\frac{1}{2} \sum_{i\neq j} \hat H_{ij} \; ,
\ee
in which the first term is the sum of the channel Hamiltonians and the
second term corresponds to cluster interactions. Modeling the Hamiltonian,
one often assumes its dependence on density and/or temperature. For example,
the effective particle spectra are often defined as functions of temperature
[62]. Therefore, in the definition of the grand Hamiltonian,
\be
\label{31}
  H = \hat H - \sum_i \mu_i \hat N_i + CV \; ,
\ee
one has to include the term $CV$ guaranteeing statistical correctness
for the approach. To this end, it is necessary to require the validity
of the conditions
\be
\label{32}
 \left\lgl \frac{\prt H}{\prt\rho_i} \right \rgl = 0 \; , \qquad
 \left\lgl \frac{\prt H}{\prt T} \right \rgl = 0 \; .
\ee
The latter reduce to the equations
\be
\label{33}
\frac{\prt C}{\prt\rho_i} = - \; \frac{1}{V} \;
\left\lgl \frac{\prt H}{\prt\rho_i} \right \rgl \; , \qquad
\frac{\prt C}{\prt T} = - \; \frac{1}{V} \;
\left\lgl \frac{\prt H}{\prt T} \right \rgl \;  ,
\ee
defining $C = C(\{\rho_i\}, T)$.

Only under the conditions of the statistical correctness (\ref{32}), the 
theory becomes self-consistent and satisfies all thermodynamic relations:
$$
P =  - \; \frac{\Om}{V} = - \; \frac{\prt\Om}{\prt V} \; \qquad
E = \frac{1}{V} \; \lgl \hat H \rgl =
T\; \frac{\prt P}{\prt T} \; - \; P + \mu_B \rho_B + \mu_S \rho_S \; ,
$$
$$
S = \frac{\prt P}{\prt T} = \frac{1}{T} \; ( E + P - \mu_B \rho_B
- \mu_S \rho_S ) \; , \qquad
\rho_i = \frac{1}{V} \; \lgl \hat N_i \rgl = \frac{\prt P}{\prt\mu_i}
$$
$$
 \rho_B = \sum_i B_i \rho_i =  \frac{\prt P}{\prt\mu_B} \; ,
\qquad
 \rho_S = \sum_i S_i \rho_i =  \frac{\prt P}{\prt\mu_S} \;.
$$
It is a common mistake, widely spread in literature, when the authors
forget about statistical correctness, because of which the obtained
results cannot be reliable.

Taking into account cluster interactions may seem to be a problem, since
there can exist various quark clusters, whose interactions are not known.
This obstacle can be avoided in the following way. Let us consider the
reaction of fusion of two clusters, say a cluster $a$ and cluster $b$,
into one cluster $i$, with all compositeness numbers larger than one,
so that there is the conservation of compositeness,
$$
z_a + z_b = z_i \;   ,
$$
and the conservation of mass,
$$
m_a + m_b + \Phi_{ab} = m_i \;  ,
$$
where $\Phi_{ab}$ is the interaction energy of two clusters. For the same
fusion, in the presence of a third cluster $j$, the mass conservation
reads as
$$
 m_a + m_b + m_j + \Phi_{ab} + \Phi_{aj} + \Phi_{bj} =
m_i + m_j + \Phi_{ij} \;  .
$$
From these relations, it follows the {\it potential scaling law}
\be
\label{34}
 \frac{\Phi_{ij}}{z_iz_j} = \frac{\Phi_{ab}}{z_az_b} \;  .
\ee
This law allows us to express all cluster interactions through one known
interaction, e.g., through the nucleon-nucleon interaction,
\be
\label{35}
  \Phi_{ij} = \frac{z_iz_j}{9} \; \Phi_{NN} \; ,
\ee
which is well known [49].

The microscopic hadron-quark-gluon mixture is stable, with deconfinement
being a sharp crossover [6,7], in good agreement with the QCD lattice
simulations [11-15]. In the case of a crossover, the deconfinement
temperature can be defined as the point where the derivatives of
observables have a maximum, which gives about 170 MeV. Of course,
considering different observables can result in slightly different
deconfinement temperatures, which is the common situation for crossovers,
where the crossover temperature is defined conditionally. Numerical
simulations [63,64] show that pion clusters survive till around $2T_c$.

\section{Static and dynamic stability}

One more problem that arises in considering the coexistence of clusters
of different types is the possibility of their spatial stratification,
when the clusters, previously uniformly mixed, separate in space into
domains containing only one kind of clusters. Below, we illustrate this
problem by considering a two-component mixture of clusters.

Let the total number of clusters be $N = N_1 + N_2$, existing in the
volume $V = V_1 + V_2$. The system can form two kinds of mixture. One
situation corresponds to a microscopic mixture, with all clusters being
uniformly intermixed in the space. And the other case is when the clusters
of each type are spatially separated into different domains, thus forming
a macroscopic mixed state. The microscopic mixture is more thermodynamically
stable when its free energy $F_{mix}$ is lower than the free energy
$F_{sep}$ of the separated state of the macroscopic mixture,
\be
\label{36}
 F_{mix} < F_{sep} \;  .
\ee
Calculating the free energy in the correlated mean-field approximation,
we use the notation for the mean interaction intensity
\be
\label{37}
 \Phi_{ij} = \int V_{ij}(\br) g_{ij}(\br)\; d\br \;  ,
\ee
in which $V_{ij}({\bf r})$ is a vacuum cluster interaction and
$g_{ij}({\bf r})$ is the pair correlation function. Then from Eq. (\ref{36}),
we find the condition for the stability of the microscopic mixture
\be
\label{38}
 \Phi_{12} < \sqrt{\Phi_{11}\Phi_{22} } \; + \; \frac{TV}{N_1N_2} \;
\Dlt S_{mix} \;  ,
\ee
where $\Delta S_{mix}$ is the entropy of mixing, which can be written as
\be
\label{39}
  \Dlt S_{mix} = - N_1 \ln \; \frac{N_1}{N} \; - \;
N_2 \ln \; \frac{N_2}{N} \; .
\ee

In the case of validity of the potential scaling (\ref{34}), the stability
condition (\ref{38}) reduces to the trivial requirement that the entropy of
mixing (\ref{39}) be positive, which is certainly true. Hence, under the
validity of the potential scaling, the microscopic mixture is always
more stable and there is no stratification.

The stability condition (\ref{38}) is derived for an equilibrium situation 
by comparing the thermodynamic potentials of the microscopic mixture and
the separated stratified state. In that sense, it is a static stability
condition. But there is also a dynamic stability condition requiring
that the spectrum of elementary excitations be positive [65]. Analyzing
the dynamic stability, we take into account that the components can move
with respect to each other with the velocities ${\bf v}_1$ and
${\bf v}_2$. Such a relative motion can be due to the fact that the
fireball has been formed as a result of two colliding heavy ions or
nuclei.

Studying the spectrum of collective excitations of a microscopic mixture
in the random-phase approximation, we find that the spectrum is positive,
provided that the relative velocity ${\bf v} = {\bf v_2} - {\bf v_1}$
does not exceed by the magnitude $v \equiv |{\bf v}|$ the critical value
\be
\label{40}
  v_c = \sqrt{ \frac{\rho_2}{m_2\Phi_{11}} \left ( \Phi_{11}\Phi_{22}
- \Phi_{12}^2 \right ) } \; .
\ee
If $v < v_c$, the microscopic mixture is stable. But if $v > v_c$, the
mixture stratifies into macroscopic domains containing different sorts
of clusters each. The dynamic instability, leading to the stratification,
caused by the mutual motion of components, is called the
{\it counterflow instability}.

\vskip 2mm

In conclusion, we have explained that there are three types of mixed
systems, macroscopic, mesoscopic, and microscopic. Each kind of these
mixed states is very different from others, enjoying quite different
physical properties and needing principally different theoretical
description.

If deconfinement would be a first-order phase transition, there could
arise the macroscopic mixed state, where hadron and quark-gluon phases
would be located in separate macroscopic spatial domains. However, such
a state is not stable and would disappear even before a fireball would
equilibrate. In addition, lattice QCD simulations demonstrate that
deconfinement is not a first-order transition, but a crossover. Hence,
the macroscopic mixed state has no chance to exist. Therefore the naive
picture, when one compares the mixed hadron-quark-gluon phase with a
boiling water containing gas bubbles, has nothing to do with QCD.
Theoretical predictions, based on the macroscopic mixed model, cannot
be confronted with experiment.

The real quark-hadron mixed state can be either mesoscopic or
microscopic. These states can be stable, with deconfinement being rather
a sharp crossover.

Studying a multicomponent mixture, it is necessary to check it with
respect to the stratification instability. The components, moving through
each other, can also exhibit the counterflow instability. All these
effects need to be carefully analyzed before comparing theoretical
predictions with experimental observations.

\newpage

\begin{figure}[ht]
\centerline{
\includegraphics[width=7.5cm]{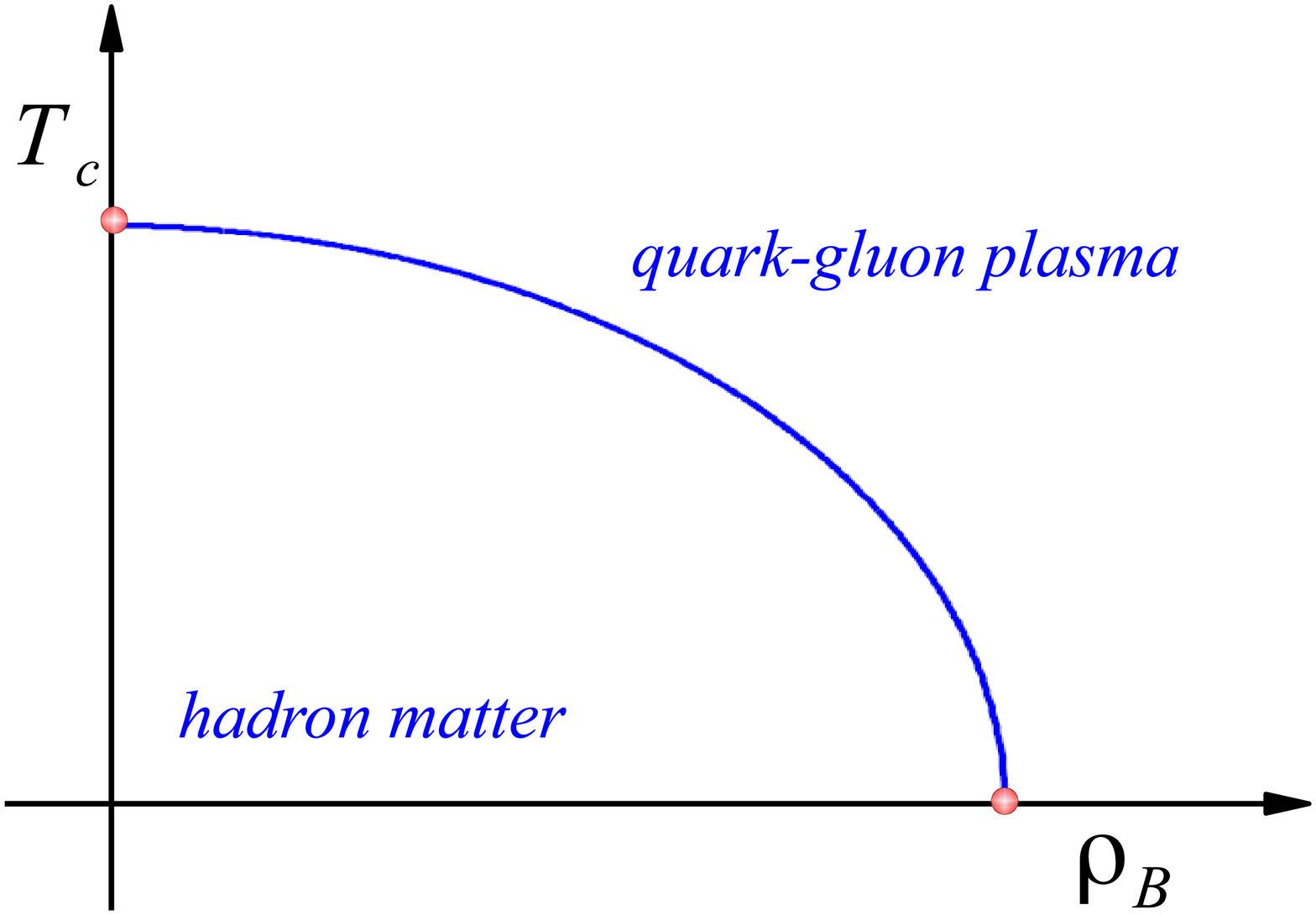} }
\caption{First-order phase-transition temperature $T_c$ as a function
of barion density.}
\label{fig:Fig.1}
\end{figure}

\vskip 2cm

\begin{figure}[ht]
\centerline{
\includegraphics[width=7.5cm]{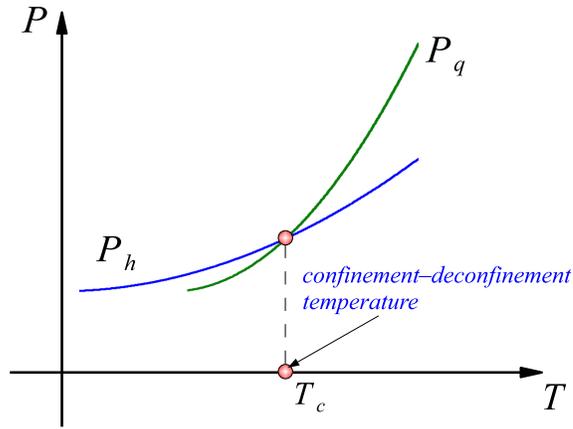} }
\caption{Pressures of the hadron matter and quark-gluon phase as functions
of temperature. }
\label{fig:Fig.2}
\end{figure}

\vskip 2cm

\begin{figure}[ht]
\centerline{
\includegraphics[width=7.5cm]{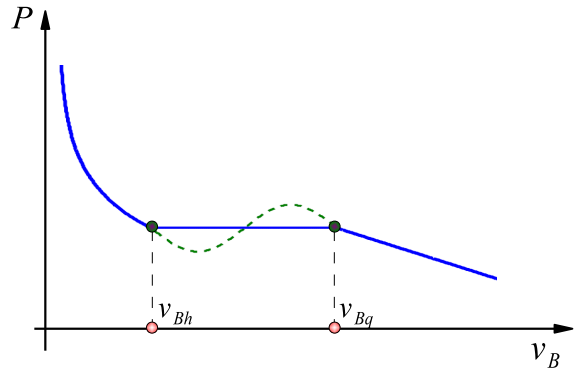} }
\caption{Maxwell construction for the pressure as a function of the reduced
barion volume.  }
\label{fig:Fig.3}
\end{figure}

\vskip 2cm

\begin{figure}[ht]
\centerline{
\includegraphics[width=7.5cm]{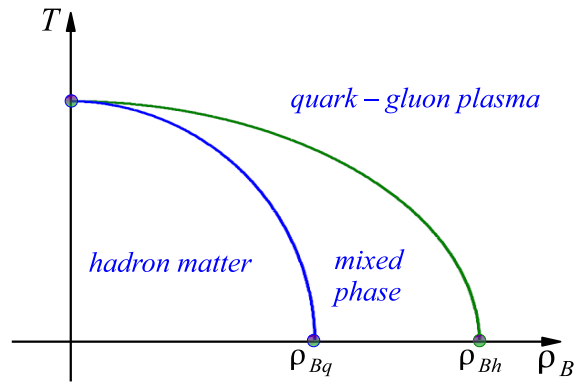} }
\caption{Mixed hadron-quark-gluon phase formally appearing around a
first-order phase transition as a result of the Maxwell construction. }
\label{fig:Fig.4}
\end{figure}

\vskip 2cm

\begin{figure}[ht]
\centerline{
\includegraphics[width=7.5cm]{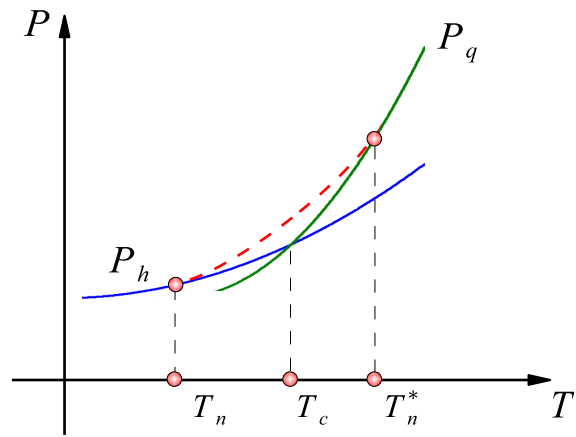} }
\caption{Smoothed pressure (dashed line) as a function of temperature,
corresponding to the Maxwell construction. }
\label{fig:Fig.5}
\end{figure}

\end{document}